\documentclass[fleqn,10pt]{wlscirep}
\usepackage[utf8]{inputenc}
\usepackage[T1]{fontenc}
\usepackage{float}
\usepackage{comment}
\usepackage{amsfonts}
\usepackage{gensymb}
\usepackage{ragged2e}
\usepackage{lineno}
\usepackage{booktabs}
\usepackage{graphicx}
\title{Histological Ageing Signatures in Routine Skin Biopsies Are Associated with Disease Burden and Mortality}

\author[1,+,*]{Kaustubh Chakradeo}
\author[2,+]{Pernille Nielsen}
\author[3, 6]{Lise Mette Rahbek Gjerdrum}
\author[3]{Gry Sahl Hansen}
\author[1]{David A Duchêne}
\author[4,1, ]{Laust H Mortensen}
\author[1,']{Majken K Jensen}
\author[1,5,'*]{Samir Bhatt}

\affil[1]{University of Copenhagen, Section of Epidemiology, Department of Public Health, Copenhagen, 1353, Denmark}

\affil[2]{Technical University of Denmark, Department of Applied Mathematics and Computer Science, Kongens Lyngby, 2800, Denmark}

\affil[3]{Department of Pathology, Copenhagen University Hospital- Zealand University Hospital, Roskilde, Denmark}

\affil[4]{Danmarks Statistik, Copenhagen, Denmark}
\affil[5]{Imperial College London, London, United Kingdom}
\affil[6]{Department of Clinical Medicine, University of Copenhagen, Copenhagen, Denmark}

\affil[*]{chakradeo@sund.ku.dk, samir.bhatt@sund.ku.dk}

\affil[+]{these authors contributed equally to this work}
\affil[']{these authors contributed equally to this work}

\begin{abstract} \label{sec:abstract}
Biological ageing is associated with progressive structural changes across multiple tissue types, yet routine histopathology remains an underexplored source of ageing-associated biological information. Here, we applied contrastive self-supervised learning to digitised skin biopsy whole-slide images from 1,787 individuals linked to Danish national health registries in order to investigate whether routine skin histology contains latent morphological signatures associated with ageing, disease burden, and mortality.
Histological feature representations extracted from unlabelled biopsy images were used to generate morphology-derived age estimates and histological age acceleration measures. These representations captured biologically interpretable tissue variation associated with epidermal atrophy, collagen disorganisation, and structural tissue degeneration. Histological ageing measures were further associated with prevalent chronic age-related diseases and long-term all-cause mortality during longitudinal follow-up.

Together, these findings suggest that routinely collected histopathology images contain clinically relevant ageing-associated information detectable using self-supervised representation learning and may provide scalable digital biomarkers for biological ageing research and computational pathology.

\end{abstract}

\begin{document}

\flushbottom
\maketitle
\thispagestyle{empty}

%\noindent Please note: Abbreviations should be introduced at the first mention in the main text – no abbreviations lists. Suggested structure of main text (not enforced) is provided below.

\section*{Highlights}\label{sec:highlights}
\begin{itemize}
    \item Self-supervised contrastive learning identified ageing-associated morphology in routine skin histology.
    \item Histological feature representations generated morphology-derived age estimates from digitised skin biopsies.
    \item Histological age acceleration was associated with age-related disease burden and long-term mortality.
    \item Dimensionality reduction and clustering analyses revealed biologically interpretable ageing-associated tissue patterns.
    \item Routine pathology archives may provide scalable digital biomarkers for biological ageing research.
\end{itemize}

\section*{Big Picture}\label{sec:bigpic}
Histopathology archives contain vast amounts of underutilised biological information collected routinely during clinical care. While skin tissue undergoes well-characterised structural changes during ageing, it remains unclear whether these changes can be systematically captured using modern artificial intelligence approaches and linked to long-term health outcomes. In this study, we applied self-supervised contrastive learning to digitised skin biopsy images linked to Danish national health registries in order to investigate whether routine histology contains latent ageing-associated morphological information.

Our findings demonstrate that morphology-derived representations extracted from routine skin histology capture biologically interpretable ageing-associated tissue variation associated with chronic disease burden and long-term mortality. These results suggest that computational pathology may enable routine archived histological material to serve as a scalable source of digital biomarkers for biological ageing and population health research. More broadly, this work highlights how self-supervised representation learning can uncover biologically meaningful structure from routinely collected clinical imaging data without requiring manual annotation.

\section*{Introduction}\label{sec:intro}
Ageing is a major driver of chronic disease, frailty, and mortality worldwide. Although chronological age is one of the strongest predictors of adverse health outcomes, individuals of the same age often exhibit substantial heterogeneity in functional decline, disease burden, and survival \cite{Kojima2019-bi}. This variability has motivated efforts to identify biomarkers that capture biological aspects of ageing beyond chronological time alone. Such biomarkers may improve understanding of ageing processes, identify individuals at elevated risk of disease, and provide clinically meaningful measures of ageing-related health status.

A broad range of molecular and physiological biomarkers have been proposed to quantify biological ageing, including telomere length, epigenetic alterations, DNA methylation patterns, mitochondrial dysfunction, and multi-omics signatures \cite{lopez2013hallmarks, lopez2023hallmarks, klemera_2006, jylhava_2017_bioagepredictors}. Among these, DNA methylation-based ageing clocks have demonstrated strong correlations with chronological age and associations with mortality and age-related disease \cite{horvath2013dna, moqri2023biomarkers}. However, no single biomarker captures the full complexity of ageing, which manifests across multiple biological scales and tissue systems \cite{sen2016epigenetic, pal2016epigenetics, aging2023biomarkers}. Moreover, many existing ageing biomarkers rely on costly molecular assays or highly specialised data collection procedures that may limit their scalability and clinical applicability.

The skin represents a potentially valuable tissue for studying human ageing. As the body's largest organ, the skin undergoes progressive structural and cellular remodelling throughout life, including epidermal thinning, collagen degradation, altered extracellular matrix organisation, and changes in cellular composition \cite{kohl2011skin, jenkins2002molecular}. Histopathological studies have demonstrated age-associated alterations in keratinocytes, fibroblasts, endothelial cells, and dermal architecture \cite{sukhovei2019difference, makrantonaki2012identification}. Because skin biopsies are routinely collected during clinical care and archived within pathology departments, they provide a scalable and underutilised resource for investigating tissue-level manifestations of ageing.

Ageing-related alterations in tissue architecture may collectively represent a form of histological ageing, whereby cumulative structural and cellular remodelling becomes detectable within routine histopathology. Such alterations include epidermal atrophy, extracellular matrix degeneration, collagen fragmentation, inflammatory disruption, and changes in dermal organisation, all of which progressively accumulate across the lifespan. Unlike molecular ageing biomarkers derived from isolated biological assays, histological ageing may capture integrated tissue-level manifestations of ageing processes occurring across multiple cellular and structural scales simultaneously. Consequently, routine histopathology images may contain latent ageing-associated information with potential relevance for disease risk and long-term health outcomes.

Recent advances in digital pathology and machine learning have enabled extraction of high-dimensional visual information from whole-slide histopathology images. In particular, self-supervised and contrastive learning approaches can learn informative image representations from large collections of unlabelled histopathology data without requiring extensive manual annotation \cite{van2021deep}. These approaches have shown promise across computational pathology tasks, including disease classification, tissue characterisation, segmentation, and prognostic modelling \cite{abhisheka2023comprehensive, ullah2023tumordetnet, fremond2023interpretable, pinckaers2019neural, long2020microscopy, ding2023large, song2023artificial, graham2023one, deng2023segment, le2020utilizing, zhu2022interpretable, chen2022fast, kanwal2023detection}. However, although age prediction from medical imaging modalities has been demonstrated across several domains, it remains unclear whether routine histopathology contains clinically meaningful histological ageing signatures associated with disease burden and mortality beyond chronological age alone.

We hypothesised that routine skin histology contains latent histological ageing signatures associated with disease burden and mortality beyond chronological age alone. To investigate this, we constructed a cohort of 1,787 digitised skin biopsy whole-slide images from age-stratified male and female participants spanning a broad age range and linked these samples to nationwide Danish health registries containing longitudinal disease and mortality data. Using contrastive representation learning, we extracted histological feature embeddings from routine skin biopsies and generated histological age estimates for each participant. We subsequently evaluated whether histological age acceleration, defined as deviations between histological and chronological age, was associated with prevalent age-related diseases and long-term survival. Finally, we investigated the biological interpretability of the extracted ageing signatures through pathological feature visualisation and unsupervised clustering of the learned histological representations.

\section*{Results}\label{sec:results}
\begin{figure}[h]
    \centering
    \includegraphics[width=\textwidth]{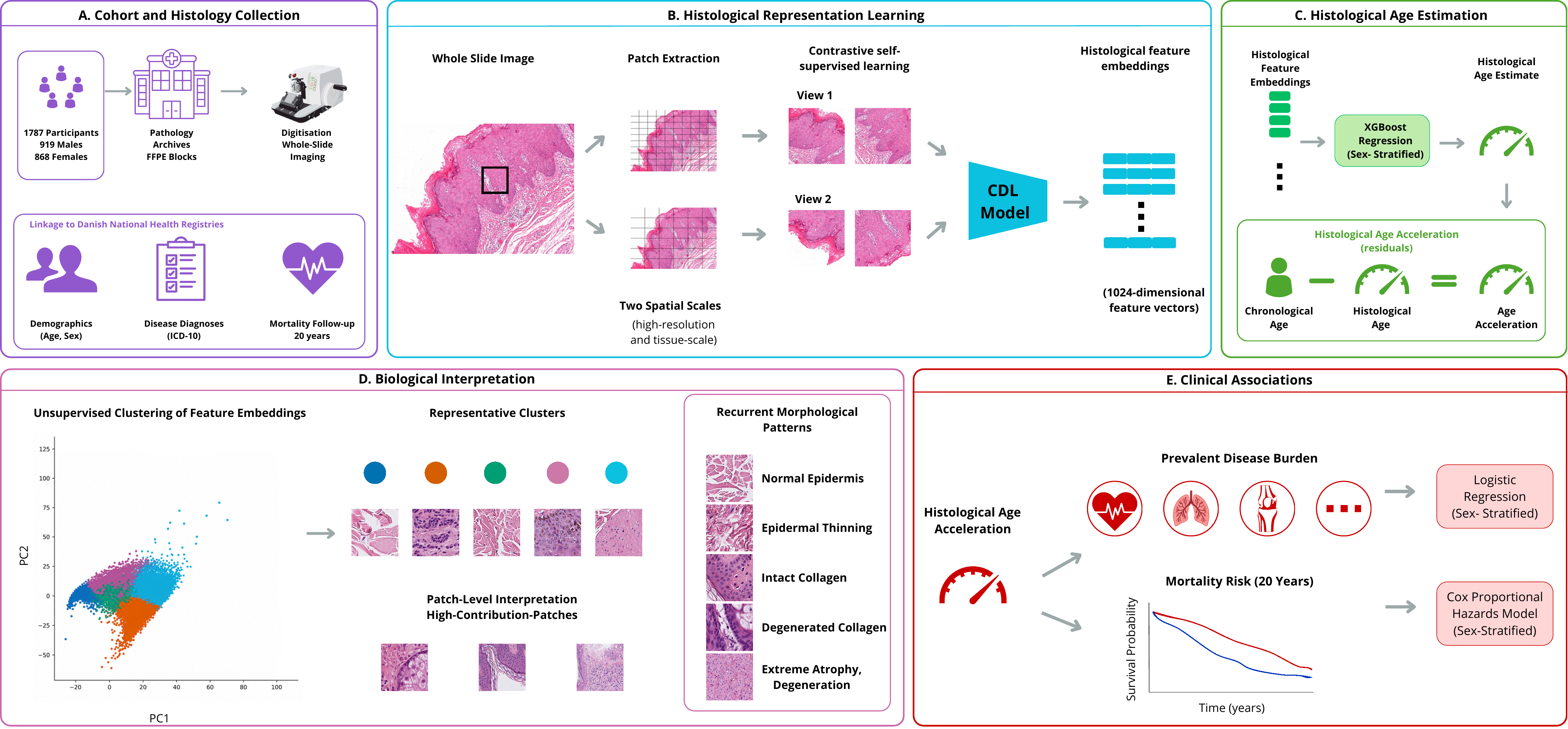}
    \caption{Overview of the study framework. 
(A) Digitised skin biopsy whole-slide images from 1,787 participants were linked to Danish national health registries containing demographic information, disease diagnoses, and long-term mortality follow-up. 
(B) Whole-slide images were divided into multiscale image patches and processed using contrastive self-supervised learning to generate histological feature embeddings capturing ageing-associated tissue morphology. 
(C) Histological feature embeddings were subsequently used to estimate histological age and derive histological age acceleration measures representing deviations from chronological age. 
(D) Biological interpretability of the learned representations was investigated using unsupervised clustering of feature embeddings and patch-level interpretation of highly weighted histological regions, revealing recurrent ageing-associated morphological patterns including epidermal thinning, collagen degeneration, and tissue atrophy. 
(E) Downstream epidemiological analyses evaluated associations between histological age acceleration, prevalent disease burden, and long-term mortality risk using sex-stratified regression and survival models.
}
    \label{fig:full}
\end{figure}

We developed a deep learning framework to extract ageing-associated histological information from digitised skin biopsy whole-slide images and investigated whether these morphology-derived signals were associated with disease burden and long-term mortality. The overall workflow is summarised in Figure \ref{fig:full}. Skin biopsy samples from 919 males and 868 females were obtained through linkage of digitised pathology archives with Danish national health registries, enabling integration of histological imaging, demographic information, disease diagnoses, and mortality follow-up over a period of up to 20 years.

Whole-slide biopsy images were divided into smaller high-resolution image patches sampled at two spatial scales in order to capture both cellular/sub-cellular morphology and broader tissue-level architecture. Contrastive self-supervised learning was subsequently used to extract histological feature embeddings from the biopsy patches without requiring manual annotation. These feature representations were then aggregated and used to generate histological age estimates for each participant. We further derived histological age acceleration measures representing deviations between histological and chronological age.

To investigate the biological and clinical relevance of the extracted histological ageing information, we evaluated associations between histological age estimates, prevalent age-related diseases, and long-term mortality. Disease classification analyses were performed using logistic regression models, while survival analyses were conducted using sex-stratified Cox proportional hazards models linked to registry-derived mortality follow-up. Finally, the biological interpretability of the extracted histological representations was examined through visualisation of highly weighted biopsy patches and unsupervised clustering of the learned feature space.

Overall, the extracted histological representations demonstrated consistent associations with chronological age, disease burden, and mortality risk. Histological age acceleration consistently improved discrimination of age-related diseases relative to chronological age alone and remained associated with elevated mortality risk after adjustment for prevalent disease burden. Visual interpretation of the learned representations further demonstrated correspondence between highly weighted histological regions and known manifestations of skin ageing, including epidermal atrophy, collagen degeneration, and altered dermal architecture.

\subsection*{Histological age estimates correlate with chronological age}

To evaluate whether routine skin histology contains ageing-associated information, histological age estimates were generated from contrastive learning feature embeddings extracted from skin biopsy whole-slide images. Whole-slide images were divided into image patches at two spatial scales in order to compare the relative contribution of cellular and sub-cellular morphology versus broader tissue-level architecture to age estimation. Feature embeddings extracted from the contrastive learning model were aggregated across patches and subsequently used to predict chronological age at the time of biopsy.

\begin{figure}[H]
    \begingroup\centering
    \includegraphics[width=\textwidth]{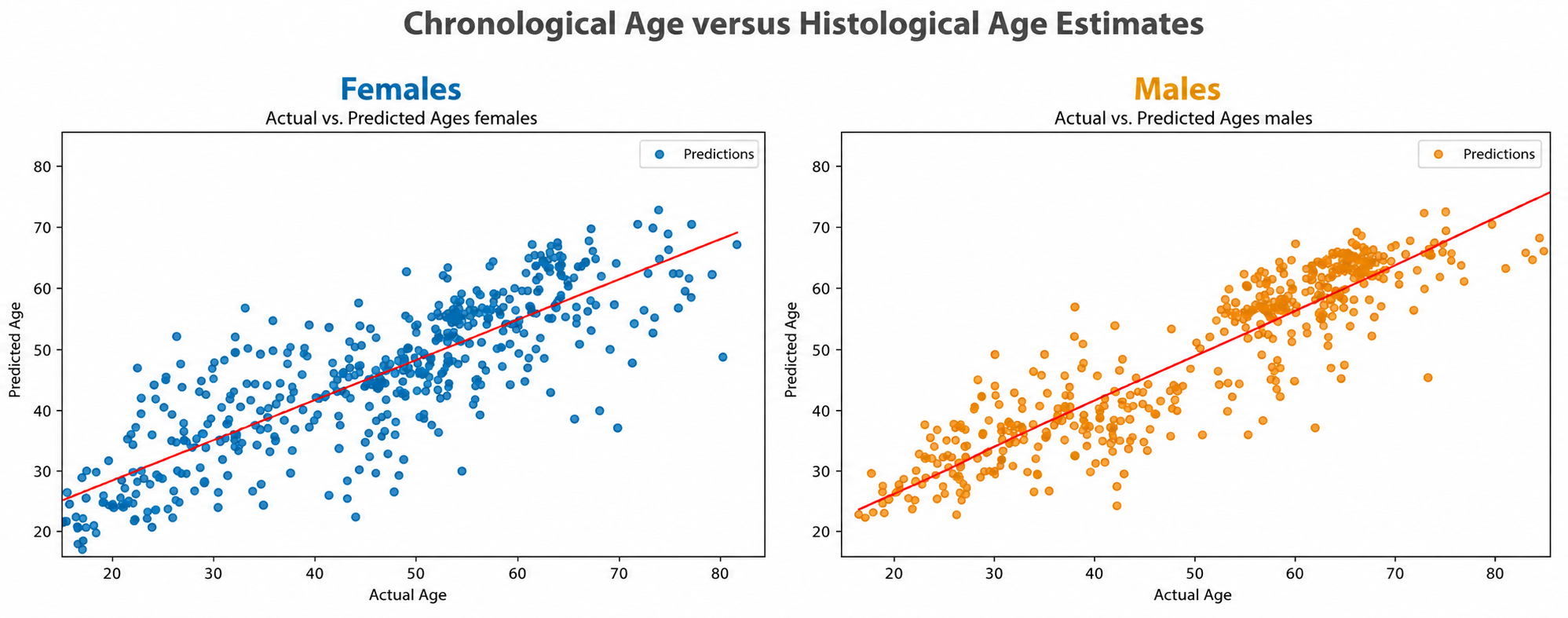}
    \caption{Relationship between chronological age and histological age estimates in female and male participants using high-resolution histological image patches. Histological age estimates demonstrated strong correlation with chronological age across both sexes, although greater variability was observed among older participants, particularly in females.}
    \label{fig:males_females}
    \endgroup
\end{figure}

Histological age estimates correlated with chronological age in both males and females (Table \ref{tab:age_result} and Figure \ref{fig:males_females}). Across nearly all age groups, smaller high-resolution image patches demonstrated lower mean absolute error (MAE) than larger tissue-scale patches, suggesting that fine-grained cellular and sub-cellular morphological patterns contain substantial ageing-associated information. In contrast, larger tissue-scale patches generally resulted in reduced predictive performance, likely reflecting loss of detailed microscopic information when broader tissue architecture was prioritised over local cellular structure.

%\centering
\begin{table}[h]
\centering
\caption{Mean absolute error (MAE) between chronological age and histological age estimates across spatial scales and age groups.}
\label{tab:age_result}
\begin{tabular}{|c|c|c|c|c|}
\hline
 Age & \# Participants  & MAE (Scale 1- Smaller) & MAE (Scale 2- Larger) & MAE (Scale 1 and 2 Combined)\\ \hline
\multicolumn{5}{|c|}{\centering \textbf{Males}} \\ \hline
$0-20$ & 100 & 1.79 (1.75 - 1.83) & 1.80 (1.76 - 1.84) &  1.80 (1.76 - 1.84)\\ \hline
$21-30$ & 100  & 2.18 (2.13 - 2.22) & 2.24 (2.19 - 2.26) & 2.14 (2.10 - 2.18)\\ \hline
$31-40$ &  100 & 2.21 (2.15 - 2.27) & 2.52 (2.44 - 2.60) & 2.45 (2.38 - 2.51) \\ \hline
$41-50$ & 100 & 2.24 (2.10 - 2.28)  & 2.75 (2.5 - 3.0)  & 2.72 (2.55 - 2.80)  \\ \hline
$51-60$ &  199 & 2.46 (2.42 - 2.50)  & 2.53  (2.47 - 2.60) & 2.52 (2.47 - 2.59)  \\ \hline
$61-70$ & 217 & 2.39 (2.35 - 2.42)  & 2.53  (2.45 - 2.60) & 2.49 (2.40 - 2.57)  \\ \hline
$\ge 71$ & 103 & 4.86 (4.74 - 4.99)  & 5.85  (5.01 - 6.69) & 5.03 (4.97 - 5.09)  \\ \hline
All ages & 919 & 2.20 (1.7 - 2.9)  & 5.63  (4.30 - 8.86) & 2.55 (1.93 - 3.17)  \\ \hline
\multicolumn{5}{|c|}{\centering \textbf{Females}} \\ \hline
$0-20$ & 99 & 2.38 (2.32 - 2.43) & 2.44 (2.38 - 2.50) &  2.41 (2.35 - 2.47) \\ \hline
$21-30$ & 99  & 2.60 (2.55 - 2.65) & 3.14 (3.06 - 3.22) & 2.65 (2.55 - 2.75) \\ \hline
$31-40$ &  100 & 2.3 (2.23 - 2.37) & 2.68 (2.60 - 2.75) & 2.43 (2.36 - 2.50) \\ \hline
$41-50$ & 99 & 2.50 (2.46 - 2.53) & 2.58 (2.52 - 2.63) & 2.51 (2.46 - 2.54) \\ \hline
$51-60$ &  246 & 2.55 (2.49 - 2.61) & 2.74 (2.68 - 2.275) & 2.72 (2.67 - 2.77) \\ \hline
$61-70$ & 153 & 2.65 (2.60 - 2.71) & 2.86 (2.76 - 2.95) & 2.69 (2.65 - 2.75) \\ \hline
$\ge 71$ & 72 & 5.49 (5.42 - 5.55) & 5.9 (5.8 - 6.1) & 5.6 (5.49 - 5.72) \\ \hline
All ages & 868 & 7.48 (6.82 - 8.16) & 10.56 (10.01 - 11.11 ) & 8.51 (8.07 - 8.95) \\ \hline
\end{tabular}
\end{table}
\justifying

The predictive advantage of smaller high-resolution patches suggests that microscopic histological alterations contribute substantially to age-associated signal extraction. Features such as epidermal atrophy, collagen degeneration, and alterations in dermal cellular architecture are likely better preserved at smaller spatial scales than in larger low-resolution tissue patches. Consequently, all subsequent analyses were performed using the smaller high-resolution patch scale.

Prediction error increased progressively in older age groups, particularly among participants over the age of 70. Younger participants generally demonstrated lower prediction variability, whereas older individuals exhibited greater divergence between histological and chronological age estimates. This increasing variability likely reflects greater heterogeneity in ageing-associated tissue morphology among older individuals relative to younger participants, whose skin architecture appeared comparatively preserved and homogeneous.

\justifying

\justifying
Sex-specific differences in prediction variability were also observed. Histological age estimates among males remained relatively tightly distributed around chronological age across most age groups, whereas female participants demonstrated greater variability at older ages. This increased heterogeneity among older females was particularly evident after approximately 50 years of age and may reflect greater variability in ageing-associated tissue remodelling and structural degeneration in later life.

\subsection*{Histological interpretation of ageing-associated tissue features}

\begin{figure}[H]
    \begingroup\centering
    \includegraphics[width = \textwidth]{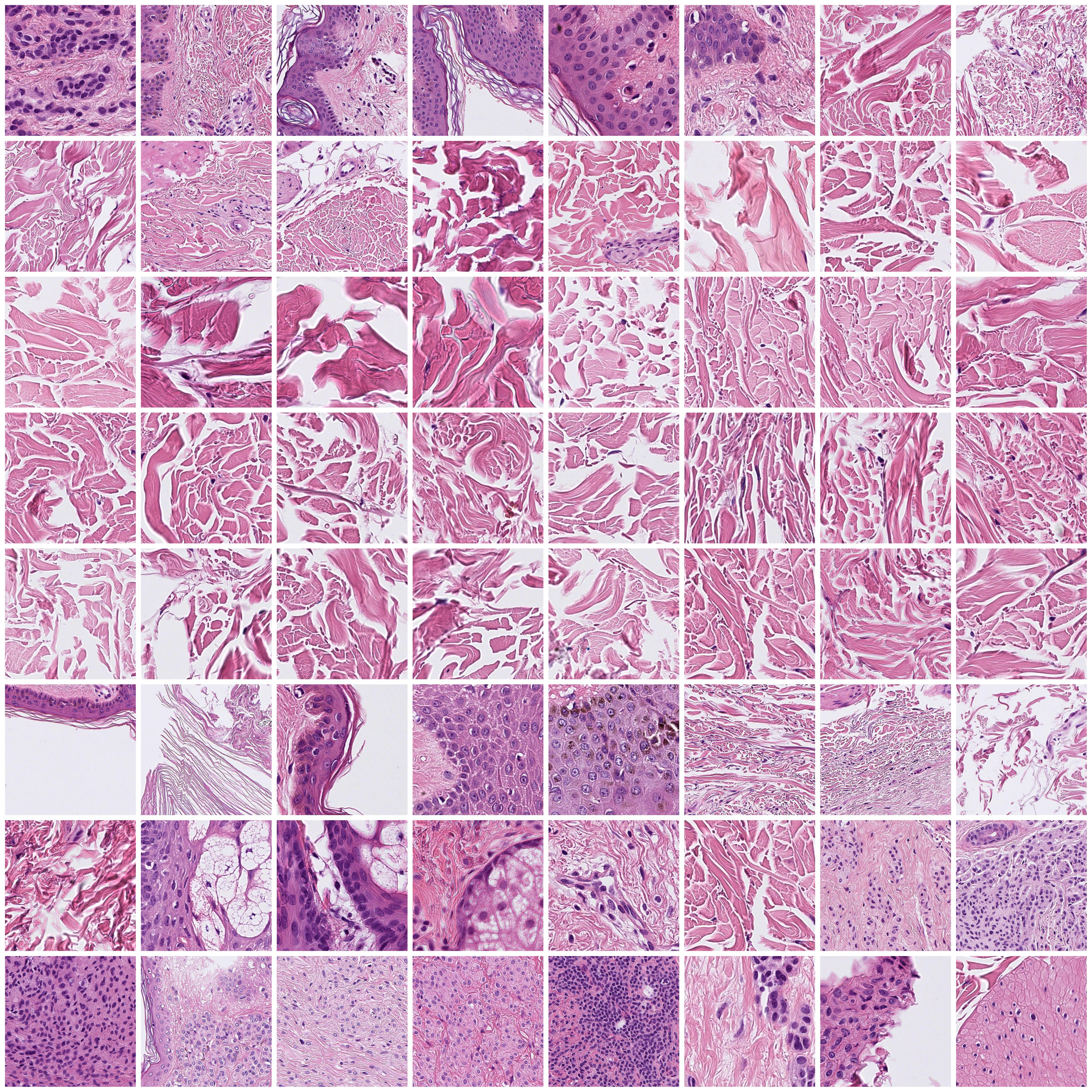}
    \caption{Representative high-contribution histological patches associated with histological age estimation. Frequently observed ageing-associated features included epidermal atrophy, collagen degradation, altered dermal architecture, inflammatory disruption, and preserved collagen organisation in younger-appearing tissue. Patches shown have undergone colour and geometric transformation for privacy preservation.}
    \label{fig:skin_samples}
    \endgroup
\end{figure}

To further investigate the histopathological features underlying the histological ageing patterns, we examined highly weighted biopsy patches contributing most strongly to histological age estimation (Figure \ref{fig:skin_samples})  \footnote{Note that for privacy reasons, the patches shown in Figure \ref{fig:skin_samples} are patches which have undergone colour and geometric transformation and may look visually different from the original patches.}. Visual inspection revealed several morphologically interpretable tissue patterns consistent with known manifestations of skin ageing, including epidermal atrophy, collagen degeneration, altered dermal architecture, and inflammatory tissue disruption.

Patches associated with younger morphological age estimates frequently demonstrated preserved epidermal structure, intact dermal architecture, and dense collagen organisation without substantial inflammatory or degenerative changes. In contrast, patches associated with older age estimates commonly exhibited epidermal thinning, fragmented collagen structure, architectural disorganisation, and inflammatory disruption. These findings are consistent with established histopathological manifestations of skin ageing \cite{branchet1990skin}. Degenerative alterations of dermal collagen represented one of the most prominent features associated with older age estimates. Younger participants frequently demonstrated preserved collagen organisation and intact dermal structure, whereas older participants showed increasingly fragmented and degraded collagen architecture. These observations are biologically plausible given the reduced fibroblast-mediated collagen production and extracellular matrix remodelling associated with ageing skin. Some highly weighted patches contained melanocytic nevi despite exclusion of biopsies with recorded skin neoplasia. Interestingly, nevi-associated patches appeared predominantly among older participants and were more frequently associated with higher histological age estimates. Although the number of such samples was limited, this observation is consistent with previous reports describing age-associated changes in nevus morphology and distribution.

Overall, patches contributing most strongly to histological age estimation consistently reflected known histopathological manifestations of skin ageing, including epidermal atrophy, collagen degeneration, and altered dermal organisation. These findings provide biological interpretability for the learned representations and support the hypothesis that the contrastive learning model captures meaningful ageing-associated tissue morphology from routine skin histology.

\subsection*{Histological feature clusters reveal ageing-associated tissue morphology}

\begin{figure}[H]
    \begingroup\centering
    \includegraphics[width = \textwidth]{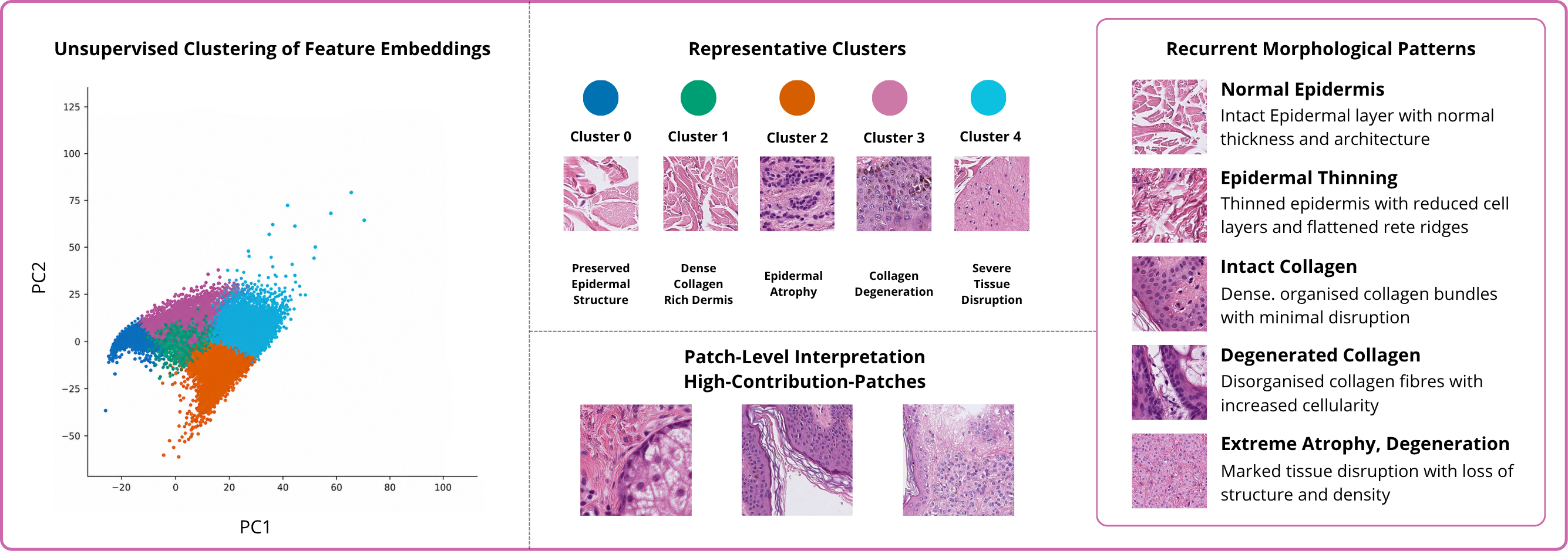}
    \caption{Principal component analysis (PCA) of morphology-derived histological feature embeddings extracted from skin biopsy whole-slide images. Distinct clusters corresponded to progressively different tissue morphologies, ranging from preserved epidermal structure and dense collagen organisation to pronounced epidermal atrophy and collagen degradation. Representative biopsy patches from each cluster are shown, illustrating ageing-associated variation in tissue architecture captured by the contrastive learning model.}
    \label{fig:pca_clusters}
    \endgroup
\end{figure}

To further investigate the interpretability of the histological ageing representations, we performed dimensionality reduction of the extracted histological feature embeddings using principal component analysis (PCA), followed by unsupervised clustering (Figure \ref{fig:pca_clusters}). The resulting feature space demonstrated distinct but partially overlapping morphological groupings, suggesting that the contrastive learning model captured continuous ageing-associated tissue variation rather than discrete pathological categories.

Visual inspection of representative biopsy patches within each cluster revealed biologically interpretable tissue characteristics associated with differing degrees of structural preservation and degeneration. Cluster 0 (blue) predominantly contained biopsies with relatively preserved epidermal architecture, organised tissue structure, and minimal visible disruption, consistent with comparatively younger or less degraded skin morphology. Cluster 1 (green) was characterised by mild epidermal atrophy and transitional tissue regions, particularly at the dermal–epidermal junction. Cluster 2 (orange) consisted primarily of dense collagen-rich dermal tissue with limited epidermal representation and comparatively preserved connective tissue organisation. Cluster 3 (pink) demonstrated features consistent with early collagen degradation and architectural disruption. Finally, Cluster 4 (cyan) exhibited pronounced epidermal atrophy, fragmented and highly degraded collagen structure, irregular tissue spacing, and a greater prevalence of extreme outlier samples. Notably, this cluster also contained samples with with marked histological age acceleration.

The progressive transition observed across clusters from preserved tissue morphology toward increasingly disorganised and degraded structural patterns is consistent with known histopathological manifestations of skin ageing. These findings provide additional biological interpretability for the learned feature representations and support the hypothesis that the model captures meaningful ageing-associated morphological variation from routine skin histology.

\subsection*{Histological age acceleration is associated with adverse health outcomes}

To investigate the clinical relevance of histological ageing information extracted from skin histology, we evaluated associations between histological age acceleration, prevalent age-related diseases, and long-term mortality. We first assessed whether histological age estimates improved discrimination of age-related disease burden relative to chronological age alone. We then examined associations between histological age acceleration and all-cause mortality using sex-stratified survival analyses linked to the Danish national registries.

\subsubsection*{Disease classification analyses}

Table \ref{tab:disease_auc} compares disease classification performance across seven age-related diseases using chronological age alone, histological age estimate, and chronological age combined with histological age acceleration. Logistic regression models were trained separately for males and females to predict prevalent disease status at the time of biopsy using diagnoses obtained from the Danish national registries.

Across all evaluated diseases and both sexes, models incorporating histological age acceleration demonstrated improved discrimination relative to models using chronological age alone. In most cases, morphology-derived histological age estimate also outperformed chronological age individually, indicating that skin histology contains ageing-associated morphological information beyond chronological age itself. The strongest performance was consistently observed when chronological age was combined with histological age acceleration, suggesting that deviations between histological and chronological age capture clinically relevant variation associated with disease burden.

The magnitude of improvement varied across diseases, with particularly pronounced gains observed for joint disease, osteoarthritis, osteoporosis, and chronic obstructive pulmonary disease. Although improvements in discrimination were generally modest, the observed effects were highly consistent across diseases and sexes despite chronological age already representing a strong baseline predictor for many age-related conditions. Together, these findings support the hypothesis that routine skin histology contains latent tissue-level signatures associated with systemic ageing processes and age-related morbidity.

\begin{table*}[h]
\centering
\caption{Comparison of disease classification performance using chronological age alone, morphology-derived histological age estimate, and chronological age combined with histological age acceleration. Values represent area under the receiver operating characteristic curve (AUC). Across diseases and sexes, models incorporating histological age acceleration consistently demonstrated improved discrimination relative to chronological age alone.}
\label{tab:disease_auc}
\begin{tabular}{|c|c|c|c|}
\hline
Disease & Actual Age & Histological Age Estimate  & Actual Age and Age Acceleration \\
\hline
\multicolumn{4}{|c|}{\textbf{Females}} \\
\hline
CVD             & 0.794 & 0.852 & \textbf{0.865} \\\hline
Cancer          & 0.723 & 0.755 & \textbf{0.771} \\\hline
Hypertension    & 0.799 & 0.833 & \textbf{0.867} \\\hline
COPD            & 0.807 & 0.847 & \textbf{0.869} \\\hline
Joint Disease   & 0.716 & 0.781 & \textbf{0.802} \\\hline
Osteoarthritis  & 0.696 & 0.744 & \textbf{0.797} \\\hline
Osteoporosis    & 0.694 & 0.724 & \textbf{0.784} \\\hline
\multicolumn{4}{|c|}{\textbf{Males}} \\
\hline
CVD             & 0.811 & 0.833 & \textbf{0.863} \\\hline
Cancer          & 0.745 & 0.757 & \textbf{0.791} \\\hline
Hypertension    & 0.755 & 0.809 & \textbf{0.820} \\\hline
COPD            & 0.763 & 0.785 & \textbf{0.798} \\\hline
Joint Disease   & 0.574 & 0.738 & \textbf{0.748} \\\hline
Osteoarthritis  & 0.719 & 0.785 & \textbf{0.791} \\\hline
Osteoporosis    & 0.782 & 0.816 & \textbf{0.847} \\\hline

\end{tabular}
\end{table*}

\subsubsection*{Survival analyses}

We next investigated whether histological ageing signals extracted from skin histology were associated with long-term mortality risk. Participants were followed for up to 20 years after biopsy through linkage to the Danish national registries. Sex-stratified Cox proportional hazards models were used to evaluate associations between chronological age, histological age acceleration, and all-cause mortality (Table \ref{tab:cox_survival}).

Across both sexes, chronological age and histological age acceleration were independently associated with mortality risk. In minimally adjusted models, hazard ratios for histological age acceleration were 1.23 (95\% CI 1.15--1.29) for females and 1.15 (95\% CI 1.13--1.18) for males. These associations remained significant after adjustment for prevalent disease burden, with disease-adjusted hazard ratios of 1.20 (95\% CI 1.18--1.22) in females and 1.19 (95\% CI 1.15--1.23) in males. Concordance indices ranged from 0.82 to 0.84 across models, indicating strong discriminative performance.

\begin{table*}[h]
\centering
\caption{
Sex-stratified Cox proportional hazards models evaluating associations between chronological age, histological age acceleration, and all-cause mortality. Hazard ratios (HR) and 95\% confidence intervals are shown for minimal and disease-adjusted models. Histological age acceleration remained associated with elevated mortality risk after adjustment for chronological age and prevalent disease burden.
}
\label{tab:cox_survival}
\begin{tabular}{|c|c|c|c|c|}
\hline
Model & Actual Age HR & Age Acceleration HR & p Value & C-index \\
\hline
\multicolumn{5}{|c|}{\textbf{Females}} \\
\hline
Minimal 
& 1.11 (1.08 -- 1.15) 
& \textbf{1.23 (1.15 -- 1.29)} 
& <0.001 
& 0.82 \\
\hline
Disease Adjusted 
& 1.08 (1.07 -- 1.10) 
& \textbf{1.20 (1.18 -- 1.22)} 
& <0.001 
& 0.84 \\
\hline
\multicolumn{5}{|c|}{\textbf{Males}} \\
\hline
Minimal 
& 1.10 (1.05 -- 1.23) 
& \textbf{1.15 (1.13 -- 1.18)} 
& <0.001 
& 0.82 \\
\hline
Disease Adjusted 
& 1.09 (1.08 -- 1.11) 
& \textbf{1.19 (1.15 -- 1.23)} 
& <0.001 
& 0.84 \\
\hline
\end{tabular}
\end{table*}

These findings suggest that deviations between histological and chronological age capture clinically relevant heterogeneity in ageing trajectories associated with mortality risk. Importantly, the persistence of these associations after adjustment for prevalent disease burden indicates that the morphology-derived histological ageing signal is not solely explained by existing age-related disease at the time of biopsy.

The observed mortality associations further support the biological relevance of the ageing-associated morphological patterns identified by the contrastive learning model. Together with the disease classification analyses and pathological clustering results, these findings indicate that routine skin histology contains latent tissue-level signatures associated with systemic ageing processes and long-term health outcomes.

\section*{Discussion}\label{sec:discussion}
In this study, we demonstrate that routine skin histology contains ageing-associated tissue signatures associated with chronological age, prevalent disease burden, and long-term mortality. Using contrastive representation learning applied to digitised skin biopsy whole-slide images, we derived histological age estimates that captured biologically interpretable tissue patterns linked to epidermal atrophy, collagen degeneration, and altered dermal organisation. Importantly, deviations between histological and chronological age, quantified as histological age acceleration, were consistently associated with adverse health outcomes across multiple disease categories and remained associated with elevated mortality risk after adjustment for prevalent disease burden.

Our findings support the concept of histological ageing, whereby cumulative structural and cellular remodelling associated with ageing becomes detectable within routine tissue histopathology. Whereas many existing ageing biomarkers rely on molecular assays or circulating biomarkers, histological ageing may capture integrated tissue-level manifestations of ageing processes occurring across multiple biological scales simultaneously. In skin tissue, these processes include epidermal thinning, extracellular matrix remodelling, collagen fragmentation, inflammatory disruption, and altered dermal organisation. The strong correspondence between highly weighted histological patches and established morphological manifestations of skin ageing provides biological interpretability for the extracted representations and suggests that the learned ageing-associated signals reflect meaningful tissue-level variation rather than arbitrary image features.

Histological age acceleration emerged as a particularly important component of the analyses. Across all evaluated diseases and both sexes, models incorporating histological age acceleration consistently demonstrated improved discrimination relative to models using chronological age alone. Although the absolute improvements in predictive performance were modest, the observed effects were highly consistent across disease categories despite chronological age already representing a strong baseline predictor for many age-related conditions. Importantly, histological age acceleration also remained associated with mortality risk after adjustment for prevalent disease burden, suggesting that deviations between histological and chronological age capture clinically relevant heterogeneity in ageing trajectories not fully explained by existing disease status alone.

The observed associations between histological ageing patterns and adverse health outcomes are biologically plausible given the systemic nature of ageing-related tissue remodelling. Ageing affects extracellular matrix integrity, cellular turnover, inflammatory regulation, fibroblast function, and vascular architecture across multiple organ systems, including the skin. As a readily accessible tissue undergoing continuous environmental and physiological exposure throughout life, the skin may therefore provide a visible and measurable manifestation of broader ageing-associated biological processes. In this context, histological age acceleration may reflect cumulative tissue damage, altered repair capacity, or heterogeneous ageing trajectories across individuals of similar chronological age.

Interpretability analyses further strengthened the biological plausibility of the extracted ageing signatures. Both highly weighted histological patches and unsupervised clustering of the learned feature space revealed progressive transitions from preserved epidermal and collagen architecture toward increasingly disorganised and degraded tissue morphology. Clusters characterised by severe collagen fragmentation, epidermal atrophy, and architectural disruption frequently corresponded to larger discrepancies between histological and chronological age estimates, consistent with marked histological age acceleration. These findings suggest that the learned feature representations capture continuous tissue-level ageing variation rather than isolated pathological entities.

An important observation was that smaller high-resolution image patches consistently outperformed larger tissue-scale patches for age estimation. This finding suggests that fine-grained cellular and sub-cellular morphology contains substantial ageing-associated information. Histological features such as epidermal atrophy, collagen degeneration, and alterations in dermal cellular architecture may therefore be better captured at smaller spatial scales than through broader low-resolution tissue architecture alone. Prediction variability also increased among older participants, particularly among older females, potentially reflecting greater heterogeneity in ageing-associated tissue remodelling later in life.

The use of contrastive self-supervised learning enabled extraction of informative histological feature representations from routinely collected pathology images without requiring extensive manual annotation. This is particularly relevant in histopathology, where large annotated datasets remain limited and expert labelling is resource intensive. Rather than relying exclusively on predefined handcrafted biomarkers, self-supervised representation learning may capture integrated tissue-level signatures reflecting cumulative structural and cellular remodelling processes associated with ageing. By learning directly from unlabelled routine pathology data, the same extracted feature space can subsequently be evaluated across multiple downstream clinical and epidemiological outcomes. Such morphology-derived representations may therefore complement existing molecular and physiological ageing biomarkers and facilitate future multimodal studies integrating histopathology, molecular profiling, and longitudinal clinical outcomes.

An important strength of this study is the use of routinely collected archival histopathology linked to nationwide longitudinal health registries. Large pathology archives represent a potentially valuable and underutilised resource for investigating tissue-level manifestations of human ageing across long follow-up periods. In settings such as Denmark, routine clinical pathology data can be linked to comprehensive registry information including disease trajectories, mortality outcomes, and demographic variables, enabling large-scale retrospective studies of ageing and disease progression. Because these tissue samples are collected as part of standard clinical care, their reuse for computational pathology research provides a scalable and cost-effective opportunity to investigate ageing-associated biological variation within real-world populations.

Several limitations should be considered when interpreting these findings. First, all biopsies originated from individuals undergoing clinical skin biopsy procedures rather than from a healthy population-based cohort. Although biopsies with recorded skin neoplasia were excluded, participants remained clinically selected and may differ systematically from the general population. Consequently, the extracted histological ageing signatures should be interpreted within the context of clinically sampled skin tissue rather than as universal markers of systemic ageing. Second, the study was conducted using a single national registry-linked cohort from Denmark, and no external validation cohort was available. Additional validation across independent populations, pathology laboratories, and imaging platforms will therefore be necessary to establish broader generalisability.

Importantly, histological age estimates derived in this study should not be interpreted as definitive measures of biological age. Rather, they represent morphology-derived tissue signatures associated with ageing-related structural variation in skin histology. Although these signatures demonstrated consistent associations with disease burden and mortality, further work incorporating orthogonal molecular ageing biomarkers, longitudinal tissue sampling, and multimodal validation will be necessary to determine the extent to which histological ageing reflects broader systemic biological ageing processes.

Additional limitations include the absence of detailed lifestyle and environmental covariates such as smoking history, ultraviolet exposure, socioeconomic status, medication use, and other behavioural risk factors that may influence skin morphology and ageing trajectories. Similarly, the observational design precludes inference regarding causality or the biological directionality of the identified associations. Future work integrating longitudinal biopsies, molecular ageing clocks, proteomics, transcriptomics, and clinical phenotyping may help clarify the mechanisms underlying histological ageing and its relationship to systemic ageing biology.

In conclusion, our findings demonstrate that routine skin histopathology contains latent ageing-associated tissue signatures associated with disease burden and long-term mortality. Histological age acceleration derived from contrastive representation learning captured clinically relevant heterogeneity in ageing trajectories beyond chronological age alone and corresponded to biologically interpretable tissue patterns including epidermal atrophy and collagen degeneration. More broadly, these results suggest that large-scale pathology archives represent a potentially valuable and underutilised resource for studying tissue-level manifestations of human ageing and their relationship to long-term health outcomes.

\section*{Methods}\label{sec:methods}
\subsection*{Study Overview}

This study investigated whether routine skin histology contains ageing-associated morphological information linked to disease burden and long-term mortality. Digitised skin biopsy whole-slide images from 1,787 individuals were retrieved from the archives of the Department of Pathology, Zealand University Hospital, Denmark, and linked to nationwide Danish health registries containing demographic information, disease diagnoses, and mortality follow-up.

The study cohort consisted of 919 males and 868 females spanning a broad age range at the time of biopsy (Table \ref{tab:age_result}). The mean chronological age at biopsy was 48.2 years (standard deviation [SD] 18.1), including 47.3 years (SD 18.6) among males and 49.2 years (SD 17.6) among females. Biopsies were obtained between 2000 and 2015, and participants were followed longitudinally through Danish national registries until 2020. During follow-up, 223 deaths were recorded.

A contrastive self-supervised learning framework was used to extract histological feature representations from the digitised biopsy images (Fig. \ref{fig:full}). These representations were subsequently used to generate histological age estimates and histological age acceleration measures. Downstream analyses evaluated associations between histological ageing information, prevalent age-related diseases, and all-cause mortality using logistic regression and Cox proportional hazards models.

\subsection*{Selection and Retrieval of Skin Biopsies }
Skin biopsy specimens were identified through the Danish National Register of Pathology (DNRP), which stores nationwide pathology records and archival formalin-fixed paraffin-embedded (FFPE) tissue blocks. Archived FFPE blocks can subsequently be retrieved for research purposes and sectioned to produce digitised histopathology whole-slide images.

Because skin biopsies are generally obtained only in the context of clinical evaluation, routine histological samples from completely healthy skin are rarely available. To obtain tissue with minimal pathological disruption, we therefore identified biopsies containing melanocytic nevi together with adjacent non-neoplastic skin tissue suitable for histological analysis. To minimise confounding related to ultraviolet-induced photoageing, biopsies were restricted to non-sun-exposed anatomical skin regions. Eligible biopsies were identified using SNOMED topology, morphology, and procedure codes corresponding to skin tissue, melanocytic nevi, and routine dermatopathology sampling procedures between 2000 and 2015 at the Department of Pathology, Zealand University Hospital, Denmark. Biopsies with malignant pathology findings or substantial additional pathological abnormalities were excluded. Detailed SNOMED code definitions are provided in Supplementary Table S1. 

Following initial filtering, biopsies with the highest probability of containing sufficient adjacent skin tissue were prioritised based on the number of FFPE blocks and materials available per requisition. Participants were subsequently stratified by sex and age category. Approximately 100 individuals were randomly sampled within each age-sex stratum, while all eligible participants above 50 years of age were included in order to increase representation among older age groups. Retrieved FFPE blocks were sectioned, stained, and digitised for downstream computational pathology analyses.

\subsection*{Histopathology image acquisition and preprocessing}
Archived histopathology slides were digitised using a Hamamatsu NanoZoomer scanner (Hamamatsu Photonics, Japan) and stored in NanoZoomer Digital Pathology Image (NDPI) format. Whole-slide images were acquired at resolutions of 2140 and 4280 pixels per inch (ppi) and consisted of high-resolution haematoxylin and eosin (H\&E)-stained skin biopsy sections.

To facilitate computational analysis, whole-slide images were divided into overlapping image patches extracted at two spatial scales in order to capture both fine-grained cellular morphology and broader tissue-level architecture. Smaller high-resolution patches measured $512 \times 512$ and $1024 \times 1024$ pixels, while larger tissue-scale patches measured $2048 \times 2048$ and $4096 \times 4096$ pixels. Adjacent patches overlapped by 50 pixels to minimise information loss at patch boundaries, corresponding to approximate physical tissue areas ranging from $2.3\,mm^2$ to $9.6\,mm^2$. Background regions containing only slide artefact or empty whitespace were excluded using colour thresholding based on the characteristic H\&E staining distribution. All retained image patches were subsequently resized to $224 \times 224$ pixels before feature extraction and model training. The use of multiple spatial scales enabled comparison of the relative contribution of local cellular morphology versus broader tissue architecture to histological age estimation and downstream epidemiological analyses.

\subsection*{Contrastive self-supervised representation learning}
Histological feature representations were extracted from image patches using a contrastive self-supervised learning framework designed to learn morphology-associated visual representations without requiring manual annotation. The framework was trained to maximise similarity between augmented views of the same image patch while maintaining separation between representations of different image patches.

For each image patch, two independently augmented views were generated using combinations of colour and geometric transformations, including random cropping, rotation, flipping, contrast adjustment, saturation adjustment, brightness perturbation, and hue modification. Augmentations were applied dynamically during training in order to increase robustness to staining variability, orientation, and local image heterogeneity commonly encountered in histopathology images. Detailed augmentation parameters are provided in Supplementary Table S2. 

Both augmented views were processed using a VGG-based convolutional neural network encoder followed by a multilayer perceptron predictor network. The encoder generated latent feature representations from each image patch, while the predictor network learned to align representations between augmented views using a cosine similarity objective function:

\begin{equation}
    -\cos{\theta} = - \frac{A \cdot B}{\lVert A \rVert \cdot \lVert B \rVert}
\end{equation}

where $A$ and $B$ represent latent feature vectors extracted from augmented views of the same image patch. A stop-gradient strategy was applied to one augmented branch during training to stabilise representation learning and prevent representational collapse.

The contrastive learning framework was trained for 100 epochs using unlabelled image patches extracted from the skin biopsy whole-slide images. Following training, each image patch was represented as a high-dimensional latent feature embedding capturing morphology-associated histological information. These embeddings were subsequently aggregated and used for downstream histological age estimation, clustering analyses, disease classification, and survival modelling.

\subsection*{Registry linkage and clinical outcomes}
Each skin biopsy whole-slide image was linked to an individual participant through the unique Danish personal identification number available within the national registries. Through registry linkage, we obtained demographic information including chronological age at biopsy, sex, date of biopsy, and vital status at end of follow-up (31 December 2020). For deceased participants, date of death and time from biopsy to death were additionally obtained. Information on prevalent age-related diseases recorded prior to or at the time of biopsy was extracted from the Danish national health registries. Evaluated disease groups included cardiovascular disease, chronic obstructive pulmonary disease (COPD), cancer (excluding skin cancer), hypertension, osteoporosis, osteoarthritis, and joint disease. Disease definitions based on ICD-10 codes are provided in Supplementary Table S3.

Additional skin-related conditions including atopic dermatitis, psoriasis, acne, and rosacea were also identified through registry linkage. However, due to the limited number of cases for these conditions within the study population, these outcomes were not included in downstream predictive analyses.

\subsection*{Feature aggregation}
Following contrastive representation learning, each image patch was represented as a high-dimensional feature embedding. Because the number of image patches varied substantially between whole-slide images, feature embeddings were aggregated at the biopsy level prior to downstream analyses. For each whole-slide image, K-means clustering was applied to group morphologically similar image patches within the learned feature space. The number of clusters was selected using the elbow method, resulting in three clusters per biopsy. Feature embeddings belonging to the same cluster were subsequently aggregated to generate compact biopsy-level histological representations for downstream modelling.

\subsection*{Histological age estimation}
Histological age estimates were generated using the extracted histological feature embeddings as predictors and chronological age at the time of biopsy as the outcome variable. Gradient boosted decision tree regression models (XGBoost) were trained to estimate morphology-derived age from the learned histological representations. All analyses were performed separately for males and females in order to account for sex-specific differences in ageing-associated tissue morphology and disease risk.

Model hyperparameters were optimised using grid search with cross-validation based on minimisation of mean absolute error (MAE). Training was performed using early stopping with a maximum of 1000 boosting rounds in order to prevent overfitting. Age estimation models were evaluated separately using feature embeddings extracted from smaller high-resolution image patches (scale 1), larger tissue-scale image patches (scale 2), and combined multiscale feature representations. Sex was included as a covariate in the prediction models. Model performance was evaluated using mean absolute error (MAE) between chronological age and histological age estimates. Histological age acceleration was subsequently defined as the residual difference between chronological age and histological age estimate and used in downstream disease and survival analyses.

\subsection*{Interpretation of learned histological representations}
To investigate the histopathological features contributing most strongly to histological age estimation, patch-level analyses were performed using the high-resolution image patch scale (scale 1). Following training of the histological age estimation model, individual image patches were evaluated according to their contribution to age prediction performance.

For each biopsy patch, histological age predictions were generated using the trained XGBoost model. Patches were subsequently ranked according to the absolute prediction error between chronological age and histological age estimate. Patches contributing to the most accurate participant-level histological age estimates were interpreted as containing highly informative ageing-associated histological features and were selected for visual inspection and pathological interpretation (Figure \ref{fig:skin_samples}). Representative high-contribution patches were subsequently reviewed to identify recurrent morphological characteristics associated with younger and older histological age estimates, including epidermal architecture, collagen organisation, dermal structure, inflammatory changes, and other ageing-associated tissue alterations.

To further investigate the structure and biological interpretability of the learned histological feature representations, dimensionality reduction and unsupervised clustering analyses were performed on the extracted feature embeddings.

Principal component analysis (PCA) was applied to the high-dimensional histological feature embeddings in order to visualise global structure within the learned representation space. Subsequently, K-means clustering was performed to identify groups of morphologically similar image patches within the embedding space. The number of clusters was selected using the elbow method. Representative biopsy patches from each cluster were visually inspected to identify recurrent histopathological patterns associated with differing tissue morphologies and ageing-associated structural variation (Figure \ref{fig:pca_clusters}).

\subsection*{Downstream epidemiological analyses}
To evaluate the clinical relevance of morphology-derived histological ageing information extracted from skin histology, logistic regression models were used to predict prevalent age-related diseases outcomes and long-term mortality follow-up obtained through linkage with the Danish national registries.

Disease classification analyses were performed using logistic regression models to predict prevalent age-related diseases recorded prior to or at the time of biopsy, using:
\begin{enumerate}
    \item chronological age alone,
    \item histological age estimate alone, and
    \item chronological age combined with histological age acceleration.
\end{enumerate}

Histological age acceleration was defined as the residual difference between chronological age and histological age estimate. Disease classification analyses were performed separately for males and females in order to account for sex-specific differences in ageing-associated disease risk. Evaluated disease groups included cardiovascular disease, cancer, hypertension, chronic obstructive pulmonary disease (COPD), osteoarthritis, osteoporosis, and joint disease. Model performance was assessed using the area under the receiver operating characteristic curve (AUC).

Associations between histological ageing measures and all-cause mortality were subsequently evaluated using sex-stratified Cox proportional hazards models. Participants were followed from the date of biopsy until death or end of follow-up on 31 December 2020. Two Cox regression models were evaluated separately for males and females:
\begin{enumerate}
    \item a minimally adjusted model including chronological age and histological age acceleration, and
    \item a disease-adjusted model additionally including prevalent age-related diseases recorded at the time of biopsy.
\end{enumerate}

\section*{Acknowledgements} \label{sec:ack}
KC acknowledges the grant support from William Demant Fonden (24-3395). KC acknowledges the help of Mignon Hagemeier for graphical design on the manuscript, and the help of Alexandros Katsiferis for discussions on machine learning based prediction models. MKJ acknowledges the help of Rudi Westendorp who originally envisioned the study. 

SB acknowledges support from the Novo Nordisk Foundation via The Novo Nordisk Young Investigator Award (NNF20OC0059309), which also funds KC. SB acknowledges the Danish National Research Foundation (DNRF160) through the chair grant which also supports MPK. SB acknowledges support from The Eric and Wendy Schmidt Fund For Strategic Innovation via the Schmidt Polymath Award (G-22-63345). SB acknowledge funding from the MRC Centre for Global Infectious Disease Analysis (reference MR/X020258/1), funded by the UK Medical Research Council
(MRC). This UK funded award is carried out in the frame of the Global Health EDCTP3 Joint Undertaking. SB is funded by the National Institute for Health and Care Research (NIHR) Health Protection Research Unit in Modelling and Health Economics, a partnership between the UK Health Security Agency, Imperial College London and LSHTM (grant code NIHR200908). Disclaimer: ‘The views expressed are those of the author(s) and not necessarily those of the NIHR, UK Health Security Agency or the Department of Health and Social Care.’. DAD acknowledges support from the Novo Nordisk Foundation via the Emerging Data Science Investigator award (NNF23OC0084647). MKJ is funded by the Novo Nordisk Foundation Challenge Programme: Harnessing the Power of Big Data to Address the Societal Challenge of Ageing (NNF17OC0027812), which also funds PYN. LHM acknowledges partial funding from NordForsk (105545), the Novo Nordisk Foundation (NNF17OC0027594 and NNF17OC0027812) and the Villum Foundation (“Nation-Scale Social Networks”).

\section*{Author contributions statement} \label{sec:auth_cont}
The study was conceived by PYN, KC and MKJ, while LMRG and GSH designed the study sample and extracted the data. They also were responsible for collecting, selecting and digitising the biopsy samples, and creating the dataset used for the experiment. MKJ, KC and SB conceived and designed the computational study. PYJ was involved in linking the biopsy samples with Danish Registers. KC performed all of the image preprocessing tasks. KC also designed and developed the deep learning modelling, as well as the downstream prediction tasks. KC drafted the initial manuscript and supplementary with input from PYN, MKJ, LMRG and SB. KC prepared the figures. KC was also responsible for reviews and redrafts of the manuscript and the repository. SB was involved in designing the deep learning methodology and downstream tasks. All authors contributed to revising the manuscript.

\section*{Data Availability}\label{sec:data_avail}

The digitised skin biopsy whole-slide images and linked Danish registry data used in this study contain person-sensitive information and are therefore not publicly available. Data are stored and managed through Statistics Denmark. Researchers affiliated with Danish research institutions may apply for access to non-image registry data through Statistics Denmark Research Services:
\href{https://www.dst.dk/en/TilSalg/Forskningsservice}{https://www.dst.dk/en/TilSalg/Forskningsservice}. Access to digitised histopathology whole-slide images additionally requires approval from the relevant Danish Research Ethics Committees. Code used for preprocessing, contrastive representation learning, and downstream analyses is publicly available at:
\href{https://github.com/kcuses/skin_histological_age}{https://github.com/kcuses/skin\_histological\_age.git}.

%\section*{Consent Declarations}\label{sec:consent}
%\input{09_consent}

\bibliography{sample}

\section*{Supplementary}\label{sec:supplementary}

\subsection*{Supplementary Cohort Selection and Histopathology Retrieval}

Archived skin biopsy specimens were identified through the Danish National Register of Pathology (DNRP), which stores nationwide pathology records and formalin-fixed paraffin-embedded (FFPE) tissue blocks indefinitely. Archived FFPE blocks can subsequently be retrieved for research purposes and sectioned to produce digitised histopathology whole-slide images.

Eligible biopsies were selected from non-sun-exposed anatomical skin regions in order to minimise confounding related to ultraviolet-induced photoageing. Biopsies containing melanocytic nevi together with adjacent non-neoplastic skin tissue were preferentially selected because completely healthy skin is rarely biopsied during routine clinical care. Samples were identified using SNOMED topology codes corresponding to non-sun-exposed skin regions, morphology codes corresponding to melanocytic nevi, and procedure codes associated with routine dermatopathology sampling procedures between 2000 and 2015 at the Department of Pathology, Zealand University Hospital, Denmark (Supplementary Table \ref{tab:snomed_skin_tcodes}).

Biopsies with malignant pathology findings or substantial additional pathological abnormalities were excluded. Following initial filtering, biopsies with the highest probability of containing sufficient adjacent normal skin tissue were prioritised based on the number of available FFPE blocks and materials per requisition. Participants were subsequently stratified by sex and age category. Approximately 100 individuals were randomly sampled within each age-sex stratum, while all eligible participants above 50 years of age were included in order to increase representation among older age groups. Retrieved FFPE blocks were sectioned, stained, and digitised for downstream computational pathology analyses. Very few biopsies were lost due to technical difficulties, resulting in a final cohort of 1,787 participants.

\begin{table}[ht]
\centering
\caption{SNOMED codes for non-sun-exposed skin areas}
\label{tab:snomed_skin_tcodes}
\begin{tabular}{|l|l|}
\hline
SNOMED code &  \\
\hline
T01000 & Skin \\
T02403 & Skin on flank \\
T02424 & Skin on chest \\
T02430(-A,-B) & \\
T02470(-A,-B) & \\
T02471(-C,-D) & \\
T02480 & \\ 
T02487 & \\
\hline
\end{tabular}

\end{table}

\subsection*{Supplementary Contrastive Deep Learning (CDL) framework}

Contrastive self-supervised learning was used to extract histological feature representations from digitised skin biopsy whole-slide images without requiring manual annotation. The framework was designed to learn image representations by maximising similarity between augmented views of the same image patch while distinguishing representations originating from different patches.

For each biopsy image patch $V$, two augmented image views ($v_1$ and $v_2$) were generated using contrastive colour and geometric augmentations. Augmentations included random crops, brightness adjustment, contrast modification, saturation perturbation, hue perturbation, image rotations, and horizontal or vertical flipping. Unlike conventional augmentation pipelines that apply symmetric transformations, augmentations were intentionally applied asymmetrically between $v_1$ and $v_2$ in order to increase representation learning difficulty and improve robustness of the learned feature space. Augmentations were generated dynamically during training and applied with probability 0.75.

A complete overview of the augmentation strategy is provided in Supplementary Table \ref{tab:aug_views}.

\begin{table}[H]
\caption{Contrasting Augmentations to generate both views, v1 and v2, of the patch V. Some augmentations like adjusting brightness, and random crops were applied to both views, while all other augmentations were applied in a contrastive fashion for the different views. Note that all augmentations were applied with a probability of 0.75.}{}\label{tab:aug_views}%
\begin{tabular}{*{5}l}
\toprule
v1 & v2  & Augmentation Type & Probability distribution & Scale \\
\midrule
\multicolumn{5}{c}{Both views} \\
\midrule
$\checkmark$ & $\checkmark$  & Random Crop & Uniform & (224 px, 224 px) \\
$\checkmark$ & $\checkmark$  & Adjust brightness & Uniform & $\pm$ 0.75 ($\Delta 0.75$) \\
\midrule
\multicolumn{5}{c}{View 1 only} \\
\midrule
$\checkmark$ & $\times$ & Random rotation & Uniform & $\pm 90 \degree$ \\
$\checkmark$ &  $\times$ & Random vertical flip & Uniform\\
$\checkmark$ & $\times$ & Adjust contrast & Uniform & $\pm 1.9$\\
% $\checkmark$ & $\times$ & Adjust Hue ($\Delta 0.01$) \\
% $\checkmark$ & $\times$ & Adjust Hue ($\Delta 0.01$) \\
$\checkmark$ & $\times$ & Adjust saturation & Uniform & $\pm 1.1$ \\
$\checkmark$ & $\times$ & Adjust hue & Uniform & $\pm 0.01$\\
% $\checkmark$ & $\times$ & Adjust hue ($\Delta 0.01$) \\
\midrule
\multicolumn{5}{c}{View 2 only} \\
\midrule
$\times$ & $\checkmark$  & Random rotation & Uniform & $\pm 180 \degree$ \\
$\times$ & $\checkmark$  & Random horizontal flip & Uniform & \\
$\times$ & $\checkmark$  & Adjust contrast & Uniform & $\pm 2.5$\\
$\times$ & $\checkmark$  & Adjust saturation & Uniform & $\pm 0.75$ \\
$\times$ & $\checkmark$ & Adjust hue & Uniform & $\pm -0.01$ \\
\bottomrule
\end{tabular}
% \caption{}
\end{table}

\subsubsection*{Contrastive Loss function}

The contrastive representation learning framework was trained using cosine similarity loss between representations of augmented image views:

\begin{equation} \label{eq:cosine}
    -\cos{\theta} = - \frac{A \cdot B}{\lVert A \rVert \cdot \lVert B \rVert}
\end{equation}

where $A$ and $B$ represent feature embeddings corresponding to augmented image views $v_1$ and $v_2$, respectively, and $\lVert \cdot \rVert$ denotes $\ell_2$ normalisation. Cosine similarity ranges from $-1$ to $1$, with larger values indicating greater similarity between representations.

Only positive image pairs derived from the same biopsy patch were used during training. Because mini-batch sizes were substantially smaller than the total number of image patches, explicit negative pair sampling was not performed in order to reduce the probability of generating false negative pairs originating from the same whole-slide image.

\subsubsection*{Encoder and Predictor Architectures}

Augmented image views were processed using a VGG-based convolutional neural network encoder architecture. Input image patches of size $224 \times 224 \times 3$ pixels were propagated through multiple ReLU-activated convolutional blocks followed by global average pooling and fully connected layers with $\ell_2$ regularisation. The encoder generated compact high-dimensional feature embeddings representing the underlying histological morphology of each image patch.

A multilayer perceptron (MLP) predictor network was attached to the encoder architecture in order to transform representations from view $v_1$ toward representations of view $v_2$. During training, gradients were propagated only through the $v_1$ branch, while a stop-gradient operation was applied to the $v_2$ branch. This asymmetric architecture stabilised training and facilitated representation learning without requiring explicit negative image pairs.

The overall contrastive learning framework is illustrated in Supplementary Fig. \ref{fig:skin_system}.

\begin{figure}[H]
    \centering
    \includegraphics[width = \textwidth]{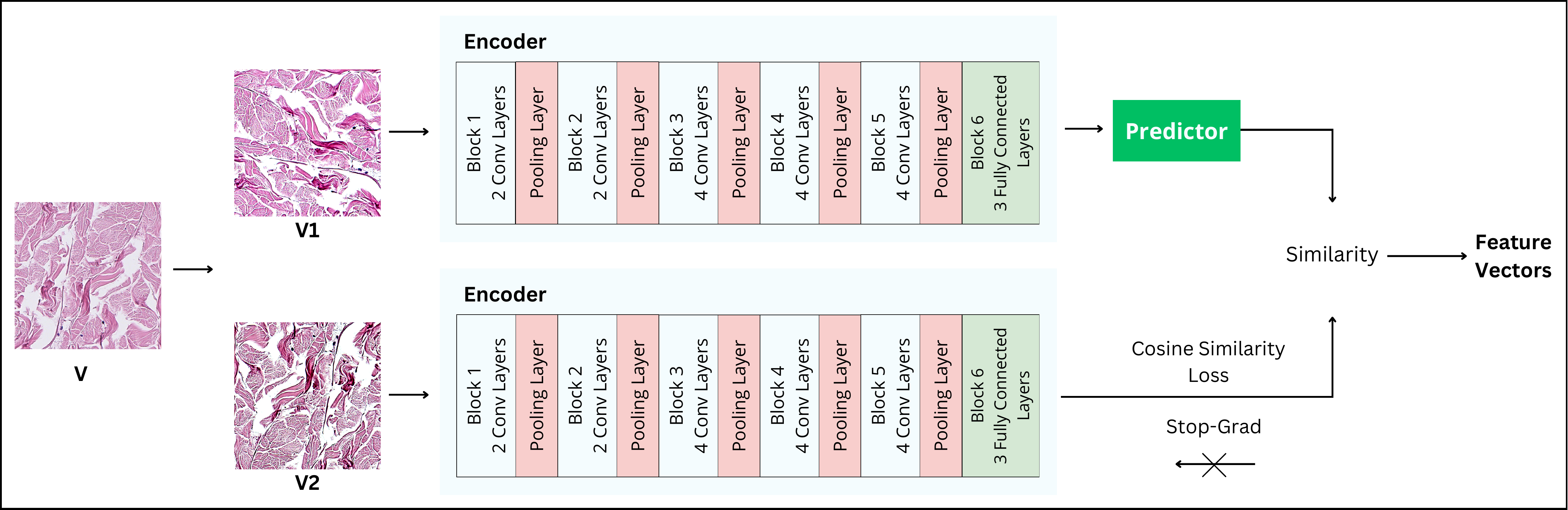}
    \caption{Contrastive Learning Method for extracting ageing information from skin biopsies}{First, contrastive augmentations are applied on the biopsy patch V to create views v1 and v2. Both v1 and v2 are passed to the encoder. The weights of v1 are updated, while on v2, a stop-grad is applied to prevent weight updation. Using the predictor MLP, the view of v1 is transformed into the view v2 using cosine similarity loss. The output of this model is the feature vector embeddings extracted from the biopsies. Thus, after training the CDL model, the representation of biopsy patch V is transformed from an image to feature vectors.}
    \label{fig:skin_system}
\end{figure}

\subsubsection*{Model training}

The contrastive representation learning framework was trained for 100 epochs using image patches extracted from the digitised skin biopsy whole-slide images. Following training, each biopsy patch was represented by a 1024-dimensional feature embedding describing its underlying histological morphology. Model pretraining was performed using an NVIDIA A6000 GPU.

\subsection*{Supplementary disease definitions}

Disease diagnoses were identified through linkage to Danish national health registries using ICD-10 diagnostic codes (Supplementary Table \ref{tab:ICD10}). Age-related diseases included cardiovascular disease, hypertension, chronic obstructive pulmonary disease (COPD), osteoporosis, osteoarthritis, joint disease, and non-skin cancer diagnoses. Additional inflammatory skin diseases were initially evaluated but were not included in downstream predictive analyses because of limited sample size.

\begin{table}[h]
    \centering
    \caption{ICD-10 codes for disease selection criteria}
    \label{tab:ICD10}
    \begin{tabular}{|c|c|}
        \hline Disease Group & ICD-10 Code \\ \hline
        Heart Disease &  I20, I21, I22, I23, I24, I25, I50, I11, I13\\
        Hypertension & I10, I11, I12, I13, I15  \\
        Joint Disease & M060, M068, M070, M071, M100, M109 \\ 
        Osteoporosis & M80, M81, M82 \\
        Osteoarthritis & M15, M16, M17, M18, M19 \\
        COPD & J40, J41, J42, J43, J44, J47, J96 \\
        Cancer & Codes below C43.0 and above C45.0\\ \hline
        Atopic Dermatitis & L20 \\
        Psoriasis & L40 \\
        Acne & L70 \\
        Rosacea & L71 \\ \hline
    \end{tabular}
    
\end{table}

%\section*{Additional information}\label{sec:appendix}
%\input{appendix}

\clearpage

\end{document}